# Greening Cloud-Enabled Big Data Storage Forensics: Syncany as a Case Study

Yee-Yang Teing, Ali Dehghantanha, *Senior Member, IEEE*, Kim-Kwang Raymond Choo, *Senior Member, IEEE*, Mohd Taufik Abdullah, and Zaiton Muda

*Abstract*—**The pervasive nature of cloud-enabled big data storage solutions introduces new challenges in the identification, collection, analysis, preservation and archiving of digital evidences. Investigation of such complex platforms to locate and recover traces of criminal activities is a time-consuming process. Hence, cyber forensics researchers are moving towards streamlining the investigation process by locating and documenting residual artefacts (evidences) of forensic value of users' activities on cloud-enabled big data platforms, in order to reduce the investigation time and resources involved in a real-world investigation. In this paper, we seek to determine the data remnants of forensic value for Syncany private cloud storage service, a popular storage engine for big data platforms. We demonstrate the types and the locations of the artefacts that can be forensically recovered. Findings from this research contribute to an in-depth understanding of cloud-enabled big data storage forensics, which can result in reduced time and resources spent in real-world investigations involve Syncany-based cloud platforms.**

*Index Terms*—**Green forensics, big data forensics, cloud forensics, Syncany forensics**

## I. INTRODUCTION

Big data and cloud computing are two information and communications technology (ICT) trends in recent years. For example, a 2013 survey by the International Data Corporation (IDC) predicted that the cloud computing spending will exceed USD 107 billion, and will drive 17% of the IT product expenditures by 2017 [1]. Gartner [2] also predicted that the cloud computing market will be worth more than USD 1.1 trillion by 2017 as more big data platforms and big data analytical solutions are deployed over cloud storage services.

There are, however, situations where public clouds are not suitable to host the big data platforms (e.g. inadequate privacy protection for data). Public cloud users are often billed for the resources consumed, including the storage input output (I/O), CPUs and memory, and the cost can become prohibitively expensive in storing, accessing and analysing high velocity, variety, veracity and volume data. There were also concerns about Cloud Service providers (CSPs) being able to infer or profile users based on the hosted / stored data [3]–[5]. Thus, it is unsurprising that cloud and big data security and privacy are current research focus [6]–[15]. Data owners in industries such as healthcare and banking are also subject to exacting regulatory requirements, which restrict the outsourcing of the data for storage and analysis. The concerns are compounded in the recent leakage of the PRISM and MUSCULAR surveillance programs that allegedly enable government and law enforcement agencies to tap into CSPs' data centres without a search warrant [16]. Therefore, in-house private dedicated cloud storage services have become an ideal solution for big data platforms.

Cloud forensics, an emerging research trend, has real-world implications for both criminal investigations and civil litigations [17]. Due to the nature of cloud-enabled big data storage solutions (e.g. data physically stored in distributed servers), identification of residual evidences may be a 'finding a needle in a haystack' exercise [18]–[21]. Even if the data could be located, traditional evidence collection tools, techniques and practices are unlikely to be adequate [22]. These issues are escalated in cross-jurisdictional investigations, which could prohibit the transfer of evidential data due to the lack of cross-nation legislative agreements in place [23]–[26]. Overcoming these investigation challenges demands significant time and resources of an investigation team [27].

In this paper, we extend our previous work [28][1] seeking to identify potential artefacts that remain on client devices and cloud servers involving the use of Syncany as a private cloud storage solution supporting big-data platforms. We attempt to answer following questions:

1. What residual artefacts can be recovered from the hard disk and memory of using Syncany desktop clients, and the

Yee-Yang Teing is with the Department of Computer Science, Faculty of Computer Science and Technology, Universiti Putra Malaysia, Serdang, 43400 Selangor, Malaysia, as well as the School of Computing, Science and Engineering, University of Salford, Salford, Greater Manchester M5 4WT, UK (e-mail: teingyeeyang@gmail.com).

Ali Dehghantanha is with the School of Computing, Science and Engineering, University of Salford, Salford, Greater Manchester M5 4WT, UK (e-mail: A. Dehghantanha@salford.ac.uk).

Kim-Kwang Raymond Choo is with the Department of Information Systems and Cyber Security, University of Texas at San Antonio, San Antonio, TX 78249-0631, USA, as well as the Information Assurance Research Group, University of South Australia, Adelaide, SA 5001, Australia (e-mail: raymond.choo@fulbrightmail.org).

Zaiton Muda and Mohd Taufik Abdullah are with the Department of Computer Science, Faculty of Computer Science and Technology, Universiti Putra Malaysia, Serdang, 43400 Selangor, Malaysia (e-mail: zaitonm@upm.edu.my and taufik@upm.edu.my).

[1] This paper is an extended conference version of [28], with more than 50% new content.

locations of the data remnants on a Windows 8.1, Ubuntu 14.04.1 LTS (Ubuntu), and Mac OS X Mavericks 10.9.5 (Mac OS) client device?

2. What data of forensics interest can be recovered from a cloud server hosting Syncany private cloud storage service and the location of the data remnants on an Ubuntu 14.04.1 LTS server?

3. What data of forensics interest can be collected from the network traffic communications between Syncany clients and servers?

We describe related works and Syncany private cloud service in the next two sections. Section IV outlines our research methodology and experiment environment setup. In Sections V and VI, we discuss findings from the clients and servers, respectively. Section VII summarises the research findings. Finally, in Section VIII, we conclude the paper and outline potential future research areas.

## II. RELATED WORK

Due to the underlying legal challenges and complexity involved in cross-jurisdictional cloud forensic investigations, Marty [29] and Shields et al. [30] proposed a proactive application-level logging mechanism, designed to log information of forensics interest which can also be used in incident response. Forensic researchers such as Dykstra and Sherman [31], Gebhardt and Reiser [32], Quick et al. [25], and Martini and Choo [33] have also presented frameworks and prototype implementations to support collection of evidential materials using Application Programming Interfaces (API) from a variety of cloud storage platforms. Although data collection using APIs could potentially reduce interactions with the CSPs, it could be limited by the APIs' features set [33].

Quick and Choo [34] and Teing et al. [35] studied the integrity of data downloaded from the web and desktop clients of Dropbox, Google Drive, Skydrive, and Symform and identified that the act of downloading data from client applications does not tamper the contents of the files, despite changes in file creation/modification times. Other cloud applications such as Evernote [36], Amazon S3 [36], Dropbox [36], [37], Google Drive [36], [38], Microsoft Skydrive [39], Amazon Cloud Drive [40], BitTorrent Sync [41], SugarSync [42], Ubuntu One [43], huBic [44], Syncany [28], Symform [35], as well as mobile cloud apps [45], [46] have also been examined. These forensics studies located artefacts of client applications from the user settings and application data resided on the media storage through keyword search. Quick and Choo [37]–[39] also identified that data erasing tools such as Eraser and CCleaner do not completely remove the data remnants from Dropbox, Google Drive, and Microsoft SkyDrive.

The first cloud forensic framework was proposed by Martini and Choo [47], which was derived based upon the frameworks of McKemmish [48] and NIST [49]. The framework was used to investigate ownCloud [50], Amazon EC2 [51], VMWare [33], and XtreemFS [52]. Quick et al. [25] further extended and validated the four-stage framework using SkyDrive, Dropbox, Google Drive, and ownCloud. Chung et al. [36] proposed a methodology for cloud investigation on Windows, Mac OS, iOS, and Android devices. The methodology was then used to investigate Amazon S3, Google Docs, and Evernote. Scanlon et al. [53] outlined a methodology for remote acquisition of evidence from decentralised file synchronisation network which utilised to investigate BitTorrent Sync [54], [55]. In another study, Teing et al. [56] proposed a methodology for investigating the newer BitTorrent Sync application (version 2.0) or any third party or Original Equipment Manufacturer (OEM) applications. Do et al. [57] proposed an adversary model for digital forensics, and they demonstrated how such an adversary model can be used to investigate mobile devices (e.g. Android smartwatch – Do et al. [58] and apps). Ab Rahman et al. [59], on the other hand, proposed a conceptual forensic-by-design framework to integrate forensics tools and best practices in the development of cloud systems. Tzuen et.al. [27], discussed about opportunities and challenges in streamlining digital investigation by referring to research that identified and documented residual artefacts.

The majority of existing published works in the field of cloud forensics have focused on client investigations on public cloud infrastructures. This paper is one of few studies which focus on both client and server forensics. Such a study may assist in reducing the carbon footprints of investigation cases involving private cloud storage solutions.

## III. BACKGROUND

Syncany is a popular open source and cross-platform cloud storage solution written in Java and supports data storage on different backend media (e.g., FTP, Box.net, WebDAV and SFTP, Amazon S3, Google Storage, IMAP, Local, Picasa and Rackspace Cloud Files). Hence, it does not require a server-side software to be installed for deployment [60]. The data synchronisation is facilitated by the sync link or repo[2] URL (similar to the concept of BitTorrent Sync, the difference being that Syncany uses a central storage infrastructure). Syncany encrypts the sync files locally (using 128-bit AES/GCM encryption) before uploading them to the central (offsite) storage (by default); hence, only clients in possession of the password can access repositories. The data model consists of three main entities as follows [60]:

- Versioning: Syncany captures different versions of a file and keeps track of the changes using metadata such as date, time, size and checksums. There are three primary versioning concepts, which are database versions, file histories, and file versions. A database version represents the point in time when the file tree is captured. Each database version contains a list of file histories, representing the identity of a file. Each file history contains a collection of file versions, representing the incarnations of a file.

- Deduplication/Chunking: Syncany uses data deduplication technique to break individual files into

---

[2] In Syncany, a repository (or commonly known as a repo) refers to a dumb storage i.e., a folder storing the packaged data and metadata about a sync folder in the backend system [60].

small chunks on the client. The chunks are represented in data blobs (each about 8-32 KB in size), which are identified by their checksum.

- Multichunking: Individual chunks are grouped into multichunks, compressed and encrypted before being uploaded to the offsite storage.

Fig. 1 shows an example of logical data model of a Syncany repository. The entities are stored locally in the form of plain-old-Java-object (POJOs) in the org.syncany.database package and tables are stored in the local HSQLDB-based database in .XML format [61].

Syncany uses a command line interface by default, but users can manually install a plugin which supports graphical user interface (GUI) as well. Syncany commands of forensic interest are as follows [62]:

- *sy init*: To initialise the repository for a new sync directory. It creates a sync-directory-specific config.xml and repo file. The former holds the local configuration information while the latter contains the chunking/crypto details required to initialise the remote repository. This command also generates the sync link in two formats: a commonly used encrypted link structured such as *syncany://storage/1/<master-salt>/<encrypted-config>*, where both <master-salt> and <encrypted-config> are base58 encoded; a plaintext link structured such as *syncany://storage/1/not-encrypted/<plaintext-config>*, where the <plaintext-config> is a base58-encoded representation of the storage/connection config.

- *sy connect*: To connect to an existing repository using the sync link or manually using the repo URL. This command is similar to *sy init*, with the difference being that it downloads the repo files from a remote storage.

- *sy status*: Lists changes made to the local sync files by comparing the local file tree (e.g., last modified dates and file sizes) with the local database.

- *sy up*: Detects changes in the local sync directory (using the 'sy status' command), indexes new files and uploads changes to the remote repository (using the 'sy up' command). File changes are packaged into new multichunks and uploaded to an offsite storage, alongside the delta metadata database.

- *sy ls-remote*: Queries the remote storage and lists the client database versions that have not yet been downloaded/processed.

- *sy down*: Detects file changes made by other clients (as identified by the 'sy ls-remote' command). The command first downloads the metadata of relevance. Then, it evaluates which multichunks are changed or required. Finally, it downloads and arranges the multichunks according to the vector clocks, when necessary.

## IV. RESEARCH METHODOLOGY

Similar to Quick and Choo [37]–[39] and Teing et al. [35], [63] Our test environments consist of four (4) VMware Workstations (VMs), one (1) for server and three (3) for desktop clients as detailed in Table I. The VMs were hosted using VMware Fusion Professional version 7.0.0 (2103067) on a Macbook Pro (Late 2012) running Mac OS X Mavericks 10.9.5, with a 2.6GHz Intel Core i7 processor and 16GB of RAM. As explained by Quick and Choo [37]–[39], virtual environments are preferred for examination of cloud applications. It is noteworthy that the Syncany application does not require an account, and hence no user account was created. The 3111[th] email messages of the Berkeley Enron email dataset (downloaded from *http://bailando.sims.berkeley.edu/enron_email.html*) were used to create a set of sample files and saved in .RTF, .TXT, .DOCX, .JPG (print screen), .ZIP, and .PDF formats, providing a basis for replication of the experiments in future. The set of sample files were placed in a new directory on the clients' workstations, before being uploaded to the servers, which subsequently being downloaded to the corresponding client devices.

In our experiments, we conducted a predefined set of activities to simulate various real world scenarios of using Syncany private cloud storage service. These included installation and uninstallation of the client applications as well as accessing, uploading, downloading, viewing, deleting, and unsyncing the sync files. A snapshot was undertaken of the VM workstations prior and after each experiment. Wireshark was deployed on the host machine to capture the communication network traffic. For the purpose of this research, we used WebDAV as the carrying protocol to enable the study of the network's inner workings. Other forensic tools prepared for experiments are highlighted in Table II. The experiments were repeated at thrice (at different dates) to ensure consistency of findings.

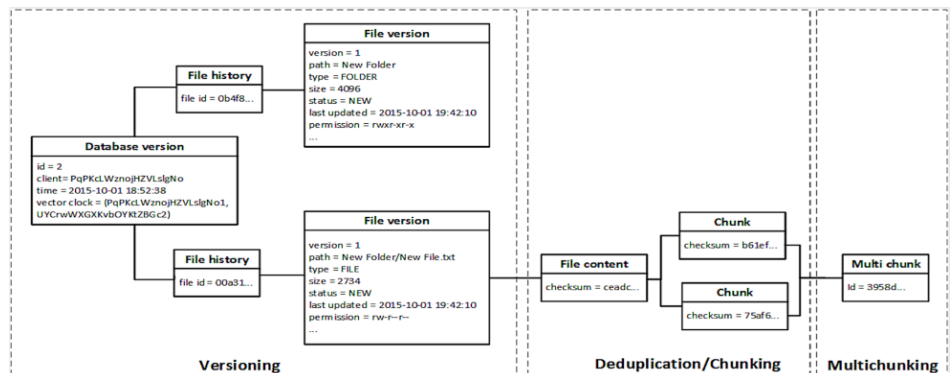

Fig. 1. Logical data model of a Syncany repository (Adapted from [12]).

TABLE II
SYSTEM CONFIGURATIONS FOR SYNCANY FORENSICS

| Server configurations | Client configurations |
|---|---|
| **Ubuntu Server**<br>Operating system : Ubuntu 14.04.1 LTS<br>Virtual memory size: 1GB<br>Virtual storage size: 20GB<br>Web application: Apache 2<br>Storage type: WebDAV<br>Database server: HSQL Database Engine 2.3.0<br>IP address/URL: http://172.16.38.180/webdav | **Windows Client**<br>Operating system: Windows 8.1 Professional (Service Pack 2, 64-bit, build 9600).<br>Virtual memory size: 2GB<br>Virtual storage size: 20GB<br>Client application: Syncany 0.4.6-alpha<br>Plugins installed: WebDAV and GUI |
| | **Ubuntu Client**<br>Operating system: Ubuntu 14.04.1 LTS<br>Virtual memory size: 1GB<br>Virtual storage size: 20GB<br>Client application: Syncany 0.4.6-alpha<br>Plugins installed: WebDAV and GUI |
| | **Mac OS Client**<br>Operating system: Mac OS X Mavericks 10.9.5<br>Virtual memory size: 1GB<br>Virtual storage size: 60GB<br>Client application: Syncany 0.4.6-alpha<br>Plugins installed: WebDAV and GUI |

TABLE I
TOOLS PREPARED FOR SYNCANY FORENSICS

| Tool | Usage |
|---|---|
| FTK Imager Version 3.2.0.0 | To create a forensic image of the .VMDK files. |
| dd version 1.3.4-1 | To produce a bit-for-bit image of the .VMEM files. |
| Autopsy 3.1.1 | To parse the file system, produce directory listings, as well as extracting/analysing the files, Windows registry, swap file/partition, and unallocated space of the forensic images. |
| raw2vmdk | To create .VMDK files from raw dd image. |
| HxD Version 1.7.7.0 | To conduct keyword searches in the unstructured datasets. |
| Volatility 2.4 | To extract the running processes and network information from the physical memory dumps, dumping files from the memory space of the Syncany client applications, and detecting the memory space of a string (using the 'pslist', 'netstat'/'netscan', 'memdump' and 'yarascan' functions). |
| SQLite Browser Version 3.4.0 | To view the contents of SQLite database files. |
| Photorec 7.0 | To data carve the unstructured datasets. |
| File juicer 4.45 | To extract files from files. |
| BrowsingHistoryView v.1.60 | To analyse the web browsing history. |
| Thumbcacheviewer Version 1.0.2.7 | To examine the Windows thumbnail cache. |
| Windows Event Viewer Version 1.0 | To view the Windows event logs. |
| Console Version 10.10 (543) | To view log files. |
| Windows File Analyser 2.6.0.0 | To analyse the Windows prefetch and link files. |
| NTFS Log Tracker | To parse and analyse the $LogFile, $MFT, and $UsnJrnl New Technology File System (NTFS) files. |
| SQL Workbench/J Build 118. | To view the contents of Hyper SQL Database (HSQLDB). |

The reliability of digital evidences rest upon the ability of the practitioners to adhere to generally accepted principles, standards, guidelines, procedures, and best practices [49], [64]. In this research, we adopted the cloud forensics framework suggested by Martini and Choo [47] for our investigation as follow:

1. *Evidence source identification and preservation.* We identified the physical hardware of interest, which contained the virtual disk (VMDK) and memory (VMEM) files in each VM folder. Then, we created a forensic copy of the VMDK and VMEM files in E01 container and raw image file (dd) formats respectively. For all the forensic images created, we calculated the MD5 and a SHA1 hash values for each original copy and subsequently verified for each working copy.

2. *Collection.* We extracted test data that matched the search terms 'syncany and 'enron' in the hard disk images. These included SQLite databases, Apple Property List (PLIST) files, event logs, shortcuts, thumbnail cache, and Windows prefetch, $MFT, $LogFile and $UsnJrnl system files. Data from the physical memory dumps were collected using Volatility, Photorec file carver, and HxD Hex Editor; network captures Wireshark and Netminer. Similar to the first stage, for all the extracted files, we calculated the MD5 and a SHA1 hash values for each original copy and subsequently verified for each working copy.

3. *Examination and analysis.* This phase is concerned with the assessment and extraction of the evidential information from the collected data. We seek to recover the following artefacts of evidential value from client devices:

- **Sync and file management metadata** – The property information generated by the Syncany client to

facilitate the synchronisation process with the hardware hosting the cloud environment. These include the node/peer IDs, usernames, device names, folder/file IDs, share keys (if any), and other property information that can facilitate file-to-device and file-to-user mappings.

- **Authentication and encryption metadata** – Artefacts describing the encryption algorithm, public/private keys or certificates, encryption salt, initialization vectors (IVs), password hashes, login URL, and (possibly) plaintext credentials that could provide the opportunity to login to the user's online account or undertake offline brute-force attacks to obtain the login password.
- **Cloud transaction logs** – Records of user actions (install, uninstall, create user account, file upload, file download, file modification, file delete, login, logoff, etc.) made to the cloud instance, which are often accompanied by the directory path and the timestamp information useful for timeline analysis.
- **Data storage** – Actual storage or temporary holding place (e.g., the application generated sync/archive folders, OS and web browser's thumbnail caches, trash folders, and unallocated space) that may be relied upon for recovering the synced files in case the cloud environment is not accessible or the actual copies have been deleted.
- **Network captures** – Copies of network packets transmitted between the client and the hardware hosting the cloud environment that contain the digital fingerprints of the user actions.
- **Volatile memory dumps** – Volatile memory is a potential source of many residual artefacts contributing to extended investigation of above mentioned items [65].

We seek to recover the following artefacts of evidential value from the server:

- **Administrative and file management metadata** – The server application generated instances that provide copies of the sync and file management as well as authentication and encryption metadata of the clients.
- **Cloud logging and authentication data:** Records of cloud transaction and authentication events/requests made by the clients to the cloud hosting instance, which are often accompanied with identifying information such as node IDs, logical addresses, session IDs, and timestamps that will facilitate provenance determination.
- **Data repositories** - The data uploaded by the users to the server (typically present in encrypted chunks or blocks), which may assist a practitioner in recovering the synced files from the cloud hosting environment.

4. *Reporting and presentation.* This final phase relates to the preparation and presentation of the information resulting from the analysis phase. As observed by Cahyani et al. [66]–[68], it is important for forensics examiners to articulate and explain complex and technical forensic terminologies to the judiciary and juries in order for them to understand "how these crimes were committed, what digital evidence is and where it may exist, and how the process of digital evidence collection was undertaken by the forensic investigators" [69].

## V. SYNCANY ANALYSIS ON DESKTOP CLIENTS

Inspection of the directory listings determined that artefacts of the Syncany client could be predominantly located in the hidden '.syncany' directory specific to the sync folders as well as user-specific 'Syncany' directory in *%Users%\<User Profile>\AppData\Roaming\*, */home/<User Profile>/.config/*, and */Users/<User Profile>/.config/* on the Windows, Ubuntu, and Mac OSX clients (respectively). The former stores folder-specific configuration while the latter manages the user-specific configuration. Both the '.syncany' and 'syncany' directories remained after uninstallation of the client applications.

### A. Sync and File Management Metadata

Analysis of the 'syncany' located the last used PID for the daemon process in the *%Syncany%/daemon.pid* file, which could be useful information for file-to-process mapping i.e., dumping the memory space using the 'memdump' function of Volatility. The */%Syncany%/daemon.xml* file held the folder metadata meant for the daemon process to identify sync directories. Each sync folder creates a 'folder' subtag in the 'folders' property to hold the directory path and information about whether the folder is enabled in the 'path' and 'enabled' entries.

Examination of the '.syncany' directories revealed 20-character machine and display names in the */%.syncany%/config.xml* file, in the 'machineName' and 'displayName' properties (respectively). The machine name is the random local machine name used to identify a computer/user for a repository, while the display name is the human readable user name for given local machine [62]. The file also described the type of backend storage in use in the 'type' entry of the 'connection' property. Knowing the type of backend storage in use may enable a practitioner to extrapolate the potential artefacts in a Syncany investigation e.g., if the WebDAV protocol is used, the web server logs can be a potential source of relevant information.

The server's displayed name was located in the */%syncany%/logs/syncany.log*, from the log entry "*1-10-15 18:52:20.274 | PluginSettingsP | main | INFO : Setting field 'username' with value 'syncanyserver'*". Further details of the sync and file management metadata could be located in the 'FILEVERSION_FULL' table of the */%.syncany%/db/local.db* database. Specifically, we recovered the unique identifications (ID), directory paths, file sizes, SHA1 checksums, and permission information (in POSIX format) for the sync folders and files specific to a repository in the 'FILEHISTORY_ID', 'PATH', 'SIZE', 'FILECONTENT_CHECKSUM', and 'POSIXPERMS' table fields. The file checksum could enable identification of a sync file outside the sync folder i.e., when the file has been moved from the sync folder.

## B. Authentication and Encryption Metadata

Our findings suggested that the files of interest for the authentication and encryption metadata are in */%syncany%/config.xml*, *%.syncany%/config.xml*, and */%.syncany%/master* files. The */%syncany%/config.xml* file records the masterkey and salt used to encrypt the sensitive information in the */%.syncany%/config.xml* file [62], in the 'configEncryptionKey' tag of the 'userConfig' property; useful when seeking to brute force the repo password stored in the 'password' tag of the 'connection' property in the */%.syncany%/config.xml* file. Meanwhile, the */%.syncany%/config.xml* file holds the plain text masterkey and salt used to derive the encryption keys for the folder-specific data, in the 'Key' and 'Salt' entries of the 'masterKey' property [62]. The master key could be located in */%.syncany%/master* file as well. We also located the URL and server's display name for the repository in the 'url' and 'username' tags, respectively.

## C. Cloud Transaction History

Inspections of the directory listings revealed the sync folder initialisation/addition time from the creation time of the '.syncany' sub-directory. Recalling the 'FILEVERSION_FULL' table of the *%.syncany%/db/local.db* database, we observed that records of the last modified and updated time could be located for the folder-specific sync directories/files in the 'LASTMODIFIED' and 'UPDATED' table fields, respectively. Sync directories/files that were modified could be differentiated from the 'VERSION' table column given the value of more than 1. Additionally, the view tables also held the directory/file modification status in the 'STATUS' table field, indicating whether the sync directories/files were newly added (new), modified (CHANGED), or deleted (deleted). We could also determine machine names for the clients that made changes (e.g., add and remove files or sub-directories) to the repositories as well as the database local times associated with the changes, in the 'DATABASEVERSION_CLIENT' and 'DATABASEVERSION_LOCALTIME' table fields, respectively (see Fig. 2). Examination of the */%syncany%/logs/syncany.log* revealed the client application's access times as well as file creation, modification, and deletion times of the sync folders/files, along with the corresponding property information. Details of the log entries of forensic

Further examination of the Windows client determined that prefetch files could be differentiated from 'SYNCANY-LATEST-X86_64-2.EXE-FDCA2B01.pf'. The prefetch files provided information about the filename and full path for the executable file, number of times the application has been loaded, last run time, and other associated timestamps. We also located shortcuts to the loader files *%Program Files (x86)%\Syncany\bin\launcher.vbs* and *%Program Files (x86)%\Syncany\unins000.exe* at *%ProgramData%\Microsoft\ Windows\Start Menu\Programs\Syncany\Syncany.lnk* and *%Pr ogramData%\Microsoft\Windows\Start Menu\Programs\Sync any\Uninstall Syncany.lnk* (respectively), providing potential for alternative methods for identifying the application installation and last run timestamps.

On the Ubuntu client, we observed that the installation created the log entry "*2015-10-01 18:32:39 install syncany:all <none> 0.4.6.alpha*" in the dpkg.log. The installation also created the log entry "*Oct 1 17:45:41 ubuntu AptDaemon.Worker: INFO: Installing local package file: /home/UbuntuPc/Desktop/syncany-latest.deb*" in the system log (Syslog). Additionally, analysis of the Zeitgeist database (*/.local/share/zeistgeist/activity.sqlite*) revealed the directory paths for the sync folders/files in the 'subj_url' and 'subj_current_uri' table fields of the 'event_view' table and original directory path in the 'subj_origin_uri' table field, which are undoubtedly useful in real world investigations. If a sync (download) file has been moved from the sync folder, then there are records remaining in the Zeitgeist database to enable identification of the sync files. We also recovered records of the types of actions made to the folders/files of relevance (e.g., Access, Create, Move, Leave, and Delete; each action creates a record in the database) in 'interpretation' table column, alongside the timestamps associated with the actions. Analysis of the Gnome's */home/<User Profile>/.local/share/recently-used.xbel* log located caches of sync directory paths accessed by the client application which included the last added, modified, and visited times in the 'bookmark' property. Fig. 3 shows that the directory path as well as last added, modified, and added times could be differentiated from the 'href', 'visited', 'modified', and 'added' entries in the 'bookmark' property. The figure also illustrates that a practitioner could identify the number of times the client application has accessed a sync folder and the application name from the 'count' and 'name' entries of the 'bookmark:applications' property.

Fig. 2. Portion of the 'FILEVERSION_FULL' table recovered in our research.

interest are in Table III.

TABLE III
ENTRIES OF PARTICULAR INTEREST IN SYNCANY.LOG

| Relevance | Examples of log entries |
|---|---|
| Assists a practitioner in the identification of the server's display name. | • 1-10-15 18:52:20.274 \| PluginSettingsP \| main \| INFO : Setting field 'username' with value 'syncanyserver' |
| Assists a practitioner in identifying the repository initiation time for a sync folder, alongside the directory path and repository URL. | • 1-10-15 18:52:27.196 \| ManagementReque \| main \| SEVE : Executing InitOperation for folder /home/suspectpc/SyncanyUbuntuClient ... <br> • 1-10-15 18:52:34.807 \| WebdavTransferM \| IntRq/SyncanyU \| INFO : WebDAV: Uploading local file /home/suspectpc/SyncanyUbuntuClient/.syncany/master to http://172.16.38.180/webdav/UbuntuRepo/master ... |
| Assists a practitioner in identifying the time when a sync folder is connected to an existing repository, including the directory path. | • 1-10-15 19:32:38.646 \| ManagementReque \| main \| SEVE : Executing ConnectOperation for folder /home/suspectpc/SyncanyWindowsDownloadToUbuntu . |
| Assists a practitioner in identifying the sync file addition time, including the property information such as the filename, file version, as well as last modified and updated times. | • 1-10-15 19:26:14.015 \| Indexer \| Thread-68 \| INFO : * Added file version: FileVersion [version=1, path=Enron3111.zip, type=FILE, status=NEW, size=30967, lastModified=Sat Dec 13 08:35:00 PST 2014, linkTarget=null, checksum=75a666ba87fef0f8425a71edcd621d0a4367aa47, updated=Thu Oct 01 19:26:14 PDT 2015, posixPermissions=rw-r--r--, dosAttributes=--a-] |
| Assists a practitioner in identifying the sync folder addition time, including the property information such as the directory name, folder version, as well as last modified and updated times. | • 1-10-15 19:42:18.751 \| FileSystemActio \| NotifyThread \| INFO : with winning version : FileVersion [version=1, path=WindowsToUbuntu, type=FOLDER, status=NEW, size=4096, lastModified=Mon Sep 28 21:40:44 PDT 2015, linkTarget=null, checksum=null, updated=Thu Oct 01 19:42:10 PDT 2015, posixPermissions=rwxr-xr-x, dosAttributes=----] |
| Enables a practitioner in determining the deletion time of a sync folder. | • 1-10-15 20:14:17.091 \| AppIndicatorTra \| PySTDIN \| INFO : Python Input Stream: Removing folder '/home/UbuntuPc/SyncanyUbuntuClient' ... |
| By searching for the 'file' tag, a practitioner can identify the filenames associated with a sync folder. | • 1-10-15 19:26:17.032 \| AppIndicatorTra \| Timer-0 \| INFO : Sending message: <updateRecentChangesGuiInternalEvent> <br>   <recentChanges> <br> <file>/home/suspectpc/SyncanyUbuntuClient/Enron3111.zip</file> <br> <file>/home/suspectpc/SyncanyUbuntuClient/Enron3111.txt</file> <br> <file>/home/suspectpc/SyncanyUbuntuClient/Enron3111.rtf</file> <br> … <br>   </recentChanges> <br> </updateRecentChangesGuiInternalEvent> |

### D. Data Storage

The sync directories could be discerned from directories containing the hidden '.syncany' sub-directory. Analysis of the thumbnail cache located thumbnail images for the synced files in *%AppData%\Local\Microsoft\Windows\Explorer\* and */home/<User Profile>/.cache/thumbnails/large/* of the Windows and Ubuntu clients (respectively). The thumbnail cache could be differentiated from the thumbnail header property 'jtEXtThumb::URI file:' that referenced the directory path for the sync folder in the latter. The files deleted (locally and remotely) could also be recovered from the unallocated space/partition.

### E. Network Analysis

Examination of the network traffic determined that only the server's IP address could be recovered. A closer inspection of the network packets revealed copies of the response body of the WebDAV protocol's PROPFIND HTTP method (i.e., "*PROPFIND /webdav/<Repo Name>/actions/ HTTP/1.1*") which contained the URLs as well as last creation and modified times of the repositories (see Fig. 4). It was also possible to recover the master key salt from the PROPFIND HTTP methods "*PUT /webdav/<Repo Name>/master HTTP/1.1*" and "*GET /webdav/<Repo Name>/master*"

- <bookmark visited="2015-10-02T01:51:56Z" modified="2015-10-02T03:16:17Z" added="2015-10-02T01:51:56Z" href="file:///home/suspectpc/SyncanyUbuntuClient">
  - <info>
    - <metadata owner="http://freedesktop.org">
        <mime:mime-type type="inode/directory"/>
      + <bookmark:applications>
      </metadata>
    </info>
  </bookmark>
- <bookmark visited="2015-10-02T02:32:15Z" modified="2015-10-02T02:32:14Z" added="2015-10-02T02:32:14Z" href="file:///home/suspectpc/SyncanyMacDownloadToUbuntu">
  - <info>
    - <metadata owner="http://freedesktop.org">
        <mime:mime-type type="inode/directory"/>
      - <bookmark:applications>
          <bookmark:application modified="2015-10-02T02:32:14Z" count="1" exec=""Syncany %u"" name="Syncany"/>
        </bookmark:applications>
      </metadata>
    </info>
  </bookmark>

Fig. 3. Portion of the recently-used.xbel log for the Syncany Ubuntu client application.

"<?xml version="1.0" encoding="utf-8"?>
<D:multistatus xmlns:D="DAV:" xmlns:ns0="DAV:">
<D:response xmlns:lp1="DAV:" xmlns:lp2="http://apache.org/dav/props/">
<D:href>/webdav/MacRepo/actions/</D:href>
<D:propstat>
<D:prop>
<lp1:resourcetype><D:collection/></lp1:resourcetype>
<lp1:creationdate>2015-09-29T15:47:30Z</lp1:creationdate>
<lp1:getlastmodified>Tue, 29 Sep 2015 15:47:30 GMT</lp1:getlastmodified>
<lp1:getetag>"1000-520e4bb450e3a"</lp1:getetag>
…".

Fig. 4. An excerpt of the response body of the PROFIND HTTP method for the Syncany repository.

*HTTP/1.1*". Since the master key salt is only created once during the repository initialisation or connection, the timestamp of which could reflect the sync folder creation or addition time. Undertaking data carving of the network traffic recovered the full repositories intact. However, it is noteworthy that the repository is encrypted by default, and hence accessing the repository may require the client application and encryption password.

### F. Memory Analysis

Examinations of the running processes using the 'pslist' function of Volatility indicated the process names, process identifiers (PID), parent process identifiers (PPID), as well as process initiation and termination times. Although the process name was masqueraded as 'java.exe', we could differentiate the PID from the last used PID recorded in the daemon.pid and syncany.log files. Examinations of the memory dumps using the 'netscan' or 'netstat' function of Volatility recovered the network information associated with the processes such as the host and server's IP addresses, port numbers, socket states, and protocols, providing an alternative method for recovery of the network information.

Undertaking data carving of the memory space of the Syncany daemon process, we recovered the files of forensic interest such as config.xml, syncany.log and local.db database intact. The presence of the data remnants in plain text also suggested that it is possible to texts from the files based on the log file entries and tag names for the configuration files. The *%.syncany%/config.xml* file could also be potentially carved from unstructured datasets, using the header and footer information of "3C 63 6F 6E 66 69 67 3E 0A…3C 2F 70 61 73 73 77 6F 72 64 3E 0A 20 20 20 3C 2F 63 6F 6E 6E 65 63 74 69 6F 6E 3E 0A 3C 2F 63 6F 6E 66 69 67 3E" in hex format or "<config>…. </connection>…</config>" in non-unicode string format, but the findings may be subject to software updates. As for the records from the 'FILEVERSION_FULL' table of the local.db database, a search for the machine name (identified from the */%.syncany%/config.xml* file) could enable future searches of the raw cell data in unstructured datasets.

### VI. Syncany Analysis on the Ubuntu Server

Examinations of the Syncany servers determined that the repositories were fully encrypted. Other than the master file

(storing the master key salt for the sync data), it appeared that accessing the sync data is only possible with the Syncany client in possession of the repository password. The repositories did not necessitate the need for the server environment, and hence it was possible to rebuild the sync folders externally (e.g., in our forensic workstation). This can be done by running the 'sy connect' command or choosing the 'Connect to an existing online storage' option on the GUI – similarly to the process of connecting to a repository from the desktop clients.

An inspection of the Apache server log determined that the log entries for the HTTP requests made by the clients to the repositories were in the format "*<Client's IP address> - <Server's name> [The time when the request was received] "<Request line>" <Status code> <Size of the repository object returned to the client> " –" "<User agent>"*". For example, we were able to determine the repository initialisation time from the log entry "*172.16.38.132 - syncanyserver [01/Oct/2015:18:52:34 -0700] "PUT /webdav/UbuntuRepo/master HTTP/1.1" 201 480 "-" "Sardine/UNAVAILABLE""*.

### VII. Reporting and Presentation

A timeline of the data from the various sources was outlined to demonstrate the cloud transactions from the clients and the servers (see Table IV). The complete timeline analysis is too broad to present fully in this paper; therefore, we only outline the timeline for the Enron3111.txt sample file from the Ubuntu client.

### VIII. Concluding Remarks

In this paper, we suggested an investigation approach for Syncany, a cloud-enabled big data platform, with the aims of reducing the required time and resources required in a real investigation. Specifically, we described the data remnants recovered from the use of Syncany private cloud storage service as a backbone for big data storage, such as config.xml, local.db, and syncany.log files on client devices. These files could provide the sync, file management, authentication, and encryption metadata essential for identifying the synced files and cloud hosting instance as well as correlating the cloud transaction records (e.g., file versioning datasets) for timelining user activities. The creation of the sync-folder-specific '.syncany' sub-directory suggested that a practitioner can identify the sync directory and the associated timestamps from the directory listing, as well as other OS-generated instances such as shortcuts, event logs, $LogFile, $MFT, $UsnJrnl, registry ('RecentDocs', 'UserAssist', 'Run', and 'ComDig32' etc.), Zeitgeist and recently-used.xbel logs, and thumbnail cache.

TABLE IV
TIMELINE FOR SYNCANY FORENSICS

| Source | Key Date | Event Type | Comment |
|---|---|---|---|
| Autopsy file List | 2015-10-01 01:14:41 | Created | Created /home/UbuntuPc/SyncanyUbuntuClient. |
| Autopsy file list | 2015-10-01 10:26:10 | Created | Created /home/UbuntuPc/SyncanyUbuntuClient/Enron3111.txt. |
| Dpkg log | 2015-10-01 18:32:39 | Created | Installed Syncany Ubuntu client application version 0.4.6.alpha - "2015-10-01 18:32:39 install syncany:all <none> 0.4.6.alpha" |
| syncany.log | 2015-10-01 18:34:12 | Last run | 1-10-15 18:34:12.064 | DaemonOperation | Thread-5 | INFO : Starting daemon operation with action RUN ... |
| syncany.log | 2015-10-01 18:52:27 | Initialised | Initialised repo for /home/suspectpc/SyncanyUbuntuClient - "1-10-15 18:52:27.196 | ManagementReque | main | SEVE : Executing InitOperation for folder /home/suspectpc/SyncanyUbuntuClient..." |
| Autopsy file list | 2015-10-01 18:52:28 | Initialised | Created repo /home/UbuntuPc/SyncanyUbuntuClient/.syncany. |
| local.db | 2015-10-01 19:26:14 | Created | Node 'UbuntuPc' (Machine Name='UYCrwWXGXKvbOYKtZBGc') added /home/UbuntuPc/SyncanyUbuntuClient/Enron3111.txt. |
| syncany.log | 2015-10-01 19:26:14 | Added | Added /home/UbuntuPc/SyncanyUbuntuClient/Enron3111.txt - "1-10-15 19:26:14.004 | Indexer | Thread-68 | INFO : * Added file version: FileVersion [version=1, path=Enron3111.txt, type=FILE, status=NEW, size=2734, lastModified=Sat Dec 13 08:33:11 PST 2014, linkTarget=null, checksum=ceadc4b7d47af4125a68e03ec3141cb3fde407ff, updated=Thu Oct 01 19:26:14 PDT 2015, posixPermissions=rw-r--r--, dosAttributes=--a-]" |
| local.db | 2015-10-01 19:42:10 | Created | Node 'WindowsPc' (Machine Name='PqPKcLWznojHZVLsIgNo') added /home/UbuntuPc/SyncanyUbuntuClient/WindowsToUbuntu. |
| local.db | 2015-10-01 19:42:10 | Created | Node 'WindowsPc' (Machine Name='PqPKcLWznojHZVLsIgNo') added /home/UbuntuPc/SyncanyUbuntuClient/WindowsToUbuntu/Enron3111.txt. |
| Autopsy file list | 2015-10-01 19:42:18 | Created | Created /home/UbuntuPc/SyncanyUbuntuClient/WindowsToUbuntu. |
| syncany.log | 2015-10-01 19:42:18 | Added | Added /home/UbuntuPc/SyncanyUbuntuClient/WindowsToUbuntu - "1-10-15 19:42:18.751 | FileSystemActio | NotifyThread | INFO : with winning version : FileVersion [version=1, path=WindowsToUbuntu, type=FOLDER, status=NEW, size=4096, lastModified=Mon Sep 28 21:40:44 PDT 2015, linkTarget=null, checksum=null, updated=Thu Oct 01 19:42:10 PDT 2015, posixPermissions=rwxr-xr-x, dosAttributes=----]" |
| syncany.log | 2015-10-01 19:42:18 | Added | Added /home/UbuntuPc/SyncanyUbuntuClient/WindowsToUbuntu/Enron3111.txt - "1-10-15 19:42:18.848 | FileSystemActio | NotifyThread | INFO : + NewFileSystemAction [file1=null, file2=FileVersion [version=1, path=WindowsToUbuntu/Enron3111.txt, type=FILE, status=NEW, size=2734, lastModified=Sat Dec 13 08:33:11 PST 2014, linkTarget=null, checksum=ceadc4b7d47af4125a68e03ec3141cb3fde407ff, updated=Thu Oct 01 19:42:10 PDT 2015, posixPermissions=rw-r--r--, dosAttributes=--a-]]" |
| Autopsy file List | 2015-10-01 19:42:18 | Created | Created /home/UbuntuPc/SyncanyUbuntuClient/WindowsToUbuntu/Enron3111.txt. |
| Autopsy file list; Trashinfo | 2015-10-01 19:48:29 | Deleted | Deleted /home/UbuntuPc/SyncanyUbuntuClient/WindowsToUbuntu/Enron3111.txt. |
| local.db | 2015-10-01 19:48:32 | Deleted | Node 'UbuntuPc' (Machine Name='UYCrwWXGXKvbOYKtZBGc') deleted /home/UbuntuPc/SyncanyUbuntuClient/WindowsToUbuntu/Enron3111.txt. |

Our examination of the physical memory indicated that the memory dumps can provide a potential alternative method for recovery of the applications cache, logs, and HTTP requests in plain text. However, the encryption password was not located (on both the hard disk and memory dumps), suggesting that a practitioner can only obtain the encryption password via an offline brute-force attack or directly from the user. Nevertheless, a practitioner must keep in mind that memory changes frequently according to user activities and will be wiped as soon as the system is shut down. Hence, obtaining a memory snapshot of a compromised system as quickly as possible increases the likelihood of obtaining the encryption key.

Analysis of the network traffic produced unique identifiable information such as URL references to the repositories and HTTP requests for the cloud transactions. We observed that the sync data were fully encrypted. This should perhaps not be surprising as Syncany fully encrypts the sync data before uploading to the offsite storage [62]. However, it is important to note that the artefacts may vary according to different network protocol implementations e.g., if the HTTPS protocol is used, then the HTTP requests will be fully encrypted.

As expected, examinations of the Syncany servers determined that the offsite repositories were fully encrypted,

TABLE V
SUMMARY OF FINDINGS FROM THE SYNCANY PRIVATE CLOUD STORAGE SERVICE

| Artefact Category | Sources of Information |
|---|---|
| Sync and file management metadata | • Sync-folder-specific *%.syncany%/config.xml* and *%.syncany%/db/local.db* <br> • User-specific *%syncany%/logs/syncany.log*, *%syncany%/daemon.xml*, and *%Syncany%/daemon.pid* |
| Authentication and encryption metadata | • Sync-folder-specific *%.syncany%/config.xml* file <br> • User-specific *%syncany%/config.xml* file |
| Cloud transaction history | • Folder-specific *%.syncany%/db/local.db* database <br> • SYNCANY-LATEST-X86_64-2.EXE-FDCA2B01.pf, Syncany.lnk, and Uninstall Syncany.lnk on the Windows client <br> • The Zeitgeist, recently-used.xbel, and system and dpkg logs of the Ubuntu client |
| Data storage | • Directories containing the '.syncany' sub-directory <br> • Copies of the thumbnail images for synced files in the thumbnail cache |
| Network analysis | • IP addresses of the client and server <br> • The timestamp information associated with the cloud transactions <br> • Copies of the clear text PROPFIND HTTP method <br> • Copies of the encrypted repositories |
| Memory analysis | • Copies of the *%.syncany%/db/local.db*, *%syncany%/logs/syncany.log*, and *%syncany%/daemon.xml* in plain text |

and it appeared that accessing the sync data is only possible with the client application and the encryption password. In the case when physical acquisition is not feasible for the server, a practitioner can facilitate the data collection with logical acquisition of the data repositories. While the machine names could be located for the nodes associated with a Syncany repository, only the web application log could provide the IP addresses of the nodes. A summary of our findings is outlined in Table V.

Collectively, our research highlighted the challenges that could arise during forensic investigations of cloud-enabled big data solutions. While the use of deduplication/chunking and encryption technologies can benefit users by providing an efficient and secure means for managing big data, evidence collection and analysis may necessitate intercepting and collecting password and utilisation of other vendor-specific applications. This can be subject to potential abuse by cyber criminals seeking to hide their tracks. Without the cooperation of the user (suspect), forensics endeavours may end up an exercise in futility. Therefore, we suggest the vendor to implement a forensically friendly logging mechanism (e.g., providing information about who accesses the data, what data has been accessed, from where did the user access the data, and when did the user access the data) that supports the collection of the raw log data outside the encrypted datasets by default. In a recent work, for example, Ab Rahman et al. [59] highlighted the importance of forensic-by-design and presented a conceptual forensic-by-design framework.

Future work would include extending this study to other private cloud storage services such as Seafile to have an up-to-date and comprehensive forensic understanding and facilitate in greening investigation of big data platforms.

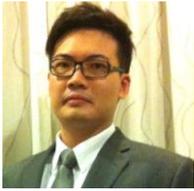
**Yee-Yang Teing** is a research fellow at the Putra University of Malaysia, and holds a Bachelor of Computer Forensics (First Class Honours). He is a Certified Ethical Hacker (CEH), Computer Hacking Forensic Investigator (CHFI) and Certified Security Analyst (ECSA). His research interests include cybercrime investigations, malware analysis and network security.

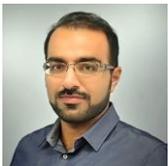
**Ali Dehghantanha** is a Marie-Curie International Incoming Fellow in Cyber Forensics and a fellow of the UK Higher Education Academy (HEA). He has served for many years in a variety of research and industrial positions. Other than Ph.D. in Cyber Security he holds many professional certificates such as GXPN, GREM, CISM, CISSP, and CCFP. He has served as an expert witness, cyber forensics analysts and malware researcher with leading players in Cyber-Security and E-Commerce.

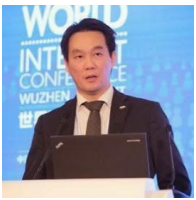

**Kim-Kwang Raymond Choo** (SM'15) received the Ph.D. in Information Security in 2006 from Queensland University of Technology, Australia. He currently holds the Cloud Technology Endowed Professorship at The University of Texas at San Antonio, and is an associate professor at University of South Australia. He has served as the Special Issue Guest Editor of ACM Transactions on Embedded Computing Systems (2016; DOI: 10.1145/3015662), ACM Transactions on Internet Technology (2016; DOI: 10.1145/3013520), Digital Investigation (2016; DOI: 10.1016/j.diin.2016.08.003), Future Generation Computer Systems (2016; DOI: 10.1016/j.future.2016.04.017), IEEE Cloud (2015; DOI: 10.1109/MCC.2015.84), IEEE Network (2016; DOI: 10.1109/MNET.2016.7764272) Journal of Computer and System Sciences (2017; DOI: 10.1016/j.jcss.2016.09.001), Multimedia Tools and Applications (2017; DOI: 10.1007/s11042-016-4081-z), Pervasive and Mobile Computing (2016; DOI: 10.1016/j.pmcj.2016.10.003), etc. He is the recipient of various awards including ESORICS 2015 Best Paper Award, Winning Team of the Germany's University of Erlangen-Nuremberg (FAU) Digital Forensics Research Challenge 2015, and 2014 Highly Commended Award by the Australia New Zealand Policing Advisory Agency, Fulbright Scholarship in 2009, 2008 Australia Day Achievement Medallion, and British Computer Society's Wilkes Award in 2008. He is a Fellow of the Australian Computer Society.